\documentclass[a4paper,fleqn,usenatbib,onecolumn]{mnras}

\usepackage{newtxtext,newtxmath}
\usepackage[T1]{fontenc}
\usepackage{ae,aecompl}
\usepackage{graphicx}
\usepackage{amsmath}
\usepackage{amssymb}

\newcommand\vx{\mathbfit{x}}
\newcommand\vk{\mathbfit{k}}
\newcommand\vs{\mathbfit{s}}
\newcommand\ve{\mathbfit{e}}
\newcommand\vn{\mathbfit{n}}
\newcommand\hP{\hat{P}}
\newcommand\hx{\hat{\mathbfit{x}}}
\newcommand\hk{\hat{\mathbfit{k}}}

\title[Power spectrum modelling of galaxy and intensity maps]{Power
  spectrum modelling of galaxy and radio intensity maps including
  observational effects}

\author[Blake et al.]{Chris Blake$^1$\thanks{E-mail:
    cblake@swin.edu.au} \\ $^1$ Centre for Astrophysics \&
  Supercomputing, Swinburne University of Technology, P.O.\ Box 218,
  Hawthorn, VIC 3122, Australia}

\date{Accepted XXX. Received YYY; in original form ZZZ}

\pubyear{2019}

\begin{document}
\label{firstpage}
\pagerange{\pageref{firstpage}--\pageref{lastpage}}
\maketitle

\begin{abstract}
Fluctuations in the large-scale structure of the Universe contain
significant information about cosmological physics, but are modulated
in survey datasets by various observational effects.  Building on
existing literature, we provide a general treatment of how fluctuation
power spectra are modified by a position-dependent selection function,
noise, weighting, smoothing, pixelization and discretization.  Our
work has relevance for the spatial power spectrum analysis of galaxy
surveys with spectroscopic or accurate photometric redshifts, and
radio intensity-mapping surveys of the sky brightness temperature
including generic noise, telescope beams and pixelization.  We
consider the auto-power spectrum of a field, the cross-power spectrum
between two fields and the multipoles of these power spectra with
respect to a curved sky, deriving the corresponding power spectrum
models, estimators, errors and optimal weights.  We note that ``FKP
weights'' for individual tracers do not in general provide the optimal
weights when measuring the cross-power spectrum.  We validate our
models using mock datasets drawn from N-body simulations\footnote{We
  provide the python code we use for these tests at
  \url{https://github.com/cblakeastro/intensitypower}.}.  Our
treatment should be useful for modelling and studying cosmological
fluctuation fields in observed and simulated datasets.
\end{abstract}

\section{Introduction}

The power spectrum of the large-scale structure of the Universe -- and
its dependence on scale, redshift and direction -- contains
significant information about the composition of the Universe and the
cosmological physics governing the growth of structure with time.
Modern cosmological surveys can trace this large-scale structure over
large volumes, by mapping the individual redshift-space positions of
galaxies or quasars, the cumulative brightness temperature of spectral
emission in a region of sky using intensity mapping in radio
wavebands, or the spectral absorption of background light by
intervening matter.

One of the central problems in cosmological analysis is to relate
these measured fluctuations in probes of large-scale structure, which
are modulated by various observational effects and analysis
approximations, to the underlying matter power spectrum which encodes
the important cosmological information.  Relevant observational
effects may include a variation in the mean background level of the
fields as a function of position (the survey selection function or
mask), noise due to the sampling of discrete objects or in the
measured brightness temperature, or smoothing of the fields in the
mapping process due to the telescope resolution.  Analysis
approximations may involve the pixelization or gridding technique
employed, and wide-angle corrections to the local plane-parallel
approximation.

Moreover, we may also utilize the cross-correlation between two
different observed fields which trace the same underlying matter
fluctuations.  Such a multi-tracer analysis offers several benefits:
(1) uncorrelated noise components in the two fields will bias the
amplitude of their auto-power spectra, but not their cross-power
spectrum; (2) an additive systematic component afflicting one of the
fields will appear in its auto-correlation but not the
cross-correlation; (3) if the fields trace a common sample variance of
matter fluctuations, such that their measurement errors are
correlated, then the noise in some joint derived parameters will be
reduced \citep{Seljak09}.

A valuable example of the ability of cross-correlation to mitigate
systematic errors arises in the joint analysis of 21-cm intensity
mapping performed by radio telescopes and galaxy redshift surveys.
Even if intensity-mapping surveys are afflicted by significant
residual components of foreground emission, cross-correlation will
allow the neutral hydrogen content of galaxies to be studied
\citep{Wolz16}.  Auto- and cross-correlation studies of current
intensity-mapping datasets, which are still limited by areal coverage,
noise and foregrounds, are presented by \citet{Chang10},
\citet{Masui13}, \citet{Wolz17} and \citet{Anderson18}.  The
scientific possibilities of radio intensity mapping will be greatly
expanded by facilities such as the Canadian Hydrogen Intensity Mapping
Experiment \citep[CHIME,][]{Bandura14}, the Hydrogen Intensity and
Real-time Analysis eXperiment \citep[HIRAX,][]{Newburgh16}, the
Tianlai Cylinder Array \citep{Xu15} and the BINGO telescope
\citep{Wuensche18}, leading up to Phase 1 of the Square Kilometre
Array \citep{Battye18}.

The imprint of observational effects in the galaxy power spectrum has
been widely modelled in the literature.  \citet{Peacock91} described
the additive and multiplicative effects of a survey mask on the
observed Fourier coefficients of the density field.  These results
were extended by \citet{FKP} who, starting from a general model of the
galaxy density field including clustering, Poisson noise and survey
selection, derived power spectrum estimators, covariance and optimal
weights.  These weights were extended by \citet{Percival04} to include
the dependence of clustering on luminosity, and by \citet{Smith15} to
encapsulate the population of halos by galaxies.  Related treatments
of the cross-power spectrum between two galaxy tracers were presented
by \citet{Smith09} and \citet{Blake13}.  \citet{Jing05} modelled the
effect on the estimated power spectrum of how fields are assigned to a
Fast Fourier Transform (FFT) grid \citep[see also,][]{Cui08} and
\citet{Sefusatti16} demonstrated the technique of interlacing to
compensate for aliasing.  Much recent work has focussed on modelling
the multipoles of the power spectrum with respect to a varying
line-of-sight direction
\citep{Yamamoto06,Beutler14,Wilson17,Beutler17,Castorina18,Blake18}.

Our work aims to review and extend these previous results by providing
a general formalism relating the 2-point statistics of fluctuations in
Fourier space to various observational effects.  This framework may be
applied to galaxy and intensity-mapping surveys, other 3D cosmological
maps, and their cross-correlation.  In particular, we extend the
literature by deriving the imprint on the auto- and cross-power
spectra of smoothing or pixelization schemes which depend on position.
Such effects are particularly relevant for radio intensity maps, which
may include a telescope beam, frequency channels and angular
pixelization across a curved sky.  We also extend the results of
\citet{FKP} to intensity mapping correlations and cross-correlations,
by considering the general optimal weighting of fields in auto- and
cross-power spectrum measurements.

Our paper is structured as follows.  In Section \ref{secmodel} we
present models for the imprint of observational effects on the
fluctuation power spectra.  After some introductory definitions
(Sections \ref{secfour}, \ref{secfluc}), we start by reviewing the
relations between the Fourier transform of the fluctuation fields and
their underlying power spectra, including the effects of a
position-dependent selection function, noise and weights, and
considering both auto- and cross-power spectra (Section \ref{secsel}).
We then derive the impact on the fluctuation power spectra if the
fields are smoothed or pixelized in a manner varying with position,
for example by a telescope beam, redshift errors, a spherical
pixelization scheme or nearest grid-point assignment (Section
\ref{secsmooth}).  We also review the effect of the discretization of
the fields onto an FFT grid (Section \ref{secdisc}).  Finally, we
summarize how the power spectra may be analysed in terms of their
multipoles with respect to a varying line-of-sight (Section
\ref{secmult}).  In Section \ref{secmeas}, we review the estimators
for the auto- and cross-power spectra and their multipoles (Section
\ref{secpkest}) and the variance in these estimators under certain
approximations, and we derive the general optimal weighting of the
fluctuation fields for measurement of these different power spectra,
providing examples for galaxy and intensity-mapping surveys and their
cross-correlation (Section \ref{secpkerr}).  In Section \ref{secsim}
we validate our models by computing the observed and predicted
multipole power spectra of mock galaxy and intensity-mapping datasets
drawn from an N-body simulation, including a variety of observational
effects.  We summarize our results in Section \ref{secconc}.

\section{Power spectrum modelling}
\label{secmodel}

\subsection{Fourier conventions}
\label{secfour}

For clarity of the subsequent derivations, we start by noting the
conventions we adopt for the Fourier transform of a function $f(\vx)$:
\begin{equation}
\tilde{f}(\vk) = \frac{1}{V} \int d^3\vx \, f(\vx) \,
e^{i\vk.\vx} , \hspace{1cm} f(\vx) = \frac{V}{(2\pi)^3} \int d^3\vk \,
\tilde{f}(\vk) \, e^{-i\vk.\vx} ,
\end{equation}
such that $f(\vx)$ and $\tilde{f}(\vk)$ have the same units, and where
$V$ is the volume of the enclosing Fourier cuboid.  We define
dimensionless Dirac delta functions $\delta_D$ in configuration and
Fourier space such that,
\begin{equation}
\tilde{\delta}_D(\vk) = \frac{1}{V} \int d^3\vx \, e^{i\vk.\vx} ,
\hspace{1cm} \delta_D(\vx) = \frac{V}{(2\pi)^3} \int d^3\vk \,
e^{i\vk.\vx} ,
\end{equation}
which are applied to functions such that,
\begin{equation}
\frac{1}{V} \int d^3\vx \, f(\vx) \, \delta_D(\vx-\vx_0) = f(\vx_0)
, \hspace{1cm} \frac{V}{(2\pi)^3} \int d^3\vk \, \tilde{f}(\vk) \,
\tilde{\delta}_D(\vk-\vk_0) = \tilde{f}(\vk_0) .
\end{equation}

\subsection{Fluctuation fields}
\label{secfluc}

We now provide some definitions related to fluctuation fields, their
correlation functions and power spectra.  Consider a function
$\delta(\vx)$ which represents the fluctuations of a field $f(\vx)$
with position $\vx$, relative to its mean ``background'' value across
many realizations of an ensemble (indicated by angled brackets) such
that,
\begin{equation}
  \delta(\vx) = f(\vx) - \langle f(\vx) \rangle ,
\label{eqdelta}
\end{equation}
and $\langle \delta(\vx) \rangle = 0$.  The field could represent the
galaxy number density distribution $f(\vx) = V \, n_g(\vx)$ (which is
dimensionless) or HI brightness temperature $f(\vx) = T_b(\vx)$ (with
dimensions of temperature).\footnote{It is appropriate to consider
  number density and temperature on the same footing, since both
  quantities do not change with the resolution of the pixelization.}
In a simple linear bias model neglecting redshift-space distortions,
the fluctuations in galaxy number density and temperature may be
described by,
\begin{equation}
    n_g(\vx) = \langle n_g(\vx) \rangle \, \left[ 1 + b_g \,
      \delta_m(\vx) \right], \hspace{1cm} T_b(\vx) = \langle T_b(\vx)
    \rangle \, \left[ 1 + b_{HI} \, \delta_m(\vx) \right] ,
\end{equation}
where $b_g$ and $b_{HI}$ are the linear bias of the galaxies and
HI-emitting objects, respectively, and $\delta_m$ is the underlying
matter overdensity.  Hence the corresponding fluctuations are:
\begin{equation}
\delta_g(\vx) = \langle n_g(\vx) \rangle \, b_g \,
\delta_m(\vx), \hspace{1cm} \delta_T(\vx) = \langle T_b(\vx) \rangle
\, b_{HI} \, \delta_m(\vx) .
\label{eqcases}
\end{equation}
The dimensionless auto-correlation function of the field between two
positions $\vx$ and $\vx'$ with separation $\vs = \vx - \vx'$,
assuming statistical homogeneity, is defined by,
\begin{equation}
\xi(\vs) = \frac{\langle f(\vx) \, f(\vx') \rangle - \langle f(\vx)
  \rangle \, \langle f(\vx') \rangle}{\langle f(\vx) \rangle \,
  \langle f(\vx') \rangle} = \frac{\langle \delta(\vx) \, \delta(\vx')
  \rangle}{\langle f(\vx) \rangle \, \langle f(\vx') \rangle} .
\label{eqcorr}
\end{equation}
Re-arranging Equation \ref{eqcorr} and adding uncorrelated noise to
the field with variance $\sigma^2(\vx)$ as a function of position, the
2-point statistics of the fluctuations can be written in the form,
\begin{equation}
  \langle \delta(\vx) \, \delta(\vx') \rangle = \langle f(\vx) \rangle
  \, \langle f(\vx') \rangle \, \xi(\vx - \vx') + \sigma^2(\vx) \,
  \delta_D(\vx-\vx') .
\label{eq2pt}
\end{equation}
Similarly, the cross-correlation of two fluctuation fields
$\delta_1(\vx) = f_1(\vx) - \langle f_1(\vx) \rangle$ and
$\delta_2(\vx) = f_2(\vx) - \langle f_2(\vx) \rangle$, assuming that
the noise in the fields is uncorrelated, is given by,
\begin{equation}
  \langle \delta_1(\vx) \, \delta_2(\vx') \rangle = \langle f_1(\vx)
  \rangle \, \langle f_2(\vx') \rangle \, \xi_c(\vx - \vx') ,
\end{equation}
in terms of the cross-correlation function $\xi_c(\vs)$.

The correlation functions of the fields may be related to their
auto-power spectra $P(\vk)$ and cross-power spectra $P_c(\vk)$ by,
\begin{equation}
  P(\vk) = \int d^3\vs \, \xi(\vs) \, e^{i\vk.\vs} , \hspace{1cm}
  P_c(\vk) = \int d^3\vs \, \xi_c(\vs) \, e^{i\vk.\vs} ,
\label{eqpk}
\end{equation}
defined here in volume units ($h^{-3}$ Mpc$^3$), including the
appropriate temperature unit for the intensity map.  As exemplified by
Equation \ref{eqcases}, our measured fields trace fluctuations in the
matter overdensity $\delta_m(\vx)$, which we model in terms of the
matter power spectrum $P_m(\vk)$.  For the purposes of this study we
assume that the redshift-space power spectra of the fields, in the
absence of any observational effects, may be described by a simple
3-parameter redshift-space distortion model \citep{Hatton98} combining
the large-scale Kaiser effect \citep{Kaiser87} imprinted by the growth
rate $f$, exponential damping from random pairwise velocities with
dispersion $\sigma_v$, and a linear bias $b$:
\begin{equation}
  P(\vk) = P(k,\mu) = \frac{(b + f \mu^2)^2 \, P_m(k)}{1 + \left( k
    \mu \sigma_v/H_0 \right)^2} , \hspace{1cm} P_c(\vk) = P_c(k,\mu) =
  \frac{(b_1 + f \mu^2) \, (b_2 + f \mu^2) \, P_m(k)}{1 + \left( k \mu
    \sigma_v/H_0 \right)^2} ,
  \label{eqpkrsd}
\end{equation}
where $\mu$ is the cosine of the angle between $\vk$ and the
line-of-sight.

In the following subsections we build a model connecting the Fourier
transform of the observed fluctuation fields to their underlying
auto-power spectra $P(\vk)$ and cross-power spectra $P_c(\vk)$, which
contains cosmological information as described by Equation
\ref{eqpkrsd}.  We include a number of practical observational and
measurement effects:
\begin{itemize}
  \item A selection function which varies with position, $\langle
    f(\vx) \rangle = \langle n_g(\vx) \rangle V$ or $\langle f(\vx)
    \rangle = \langle T_b(\vx) \rangle$,
  \item Uncorrelated noise in the field as a function of position,
    described by $\sigma^2(\vx)$ in Equation \ref{eq2pt}, including
    the specific example of Poisson noise,
  \item A weight $w(\vx)$ applied to the field to optimize the
    signal-to-noise ratio of the measurement,
  \item A smoothing function which can vary with position, with
    specific examples provided for a Gaussian telescope beam,
    frequency channels in radio observations, redshift errors, {\tt
      HEALPix}
    pixelization\footnote{\url{http://healpix.sourceforge.net}}
    \citep{Healpix05} and nearest grid point assignment,
  \item Discretization of the field onto an FFT grid.
\end{itemize}

\subsection{Relating the fluctuation fields to the power spectra}
\label{secsel}

We now develop the relationship between the observed fluctuations and
their underlying power spectra, building on existing literature.  We
can relate the observed fluctuation fields to their power spectra by
considering the Fourier transform of the weighted fields,
\begin{equation}
\tilde{\delta}(\vk) = \frac{1}{V} \int d^3\vx \, w(\vx) \, \delta(\vx)
\, e^{i\vk.\vx} ,
\label{eqdelk}
\end{equation}
where $w(\vx)$ is a general position-dependent weight\footnote{The
  weight has inverse units to those of the field, considering the
  definition presented after Equation \ref{eqdelta} -- i.e.,
  dimensionless for a galaxy survey and inverse temperature for an
  intensity map.} which may be applied to optimize the signal-to-noise
ratio of the measurement.  The average of $|\tilde{\delta}(\vk)|^2$
across realizations is,
\begin{equation}
\langle |\tilde{\delta}(\vk)|^2 \rangle = \frac{1}{V^2} \int d^3\vx
\int d^3\vx' \, w(\vx) \, w(\vx') \, \langle \delta(\vx) \,
\delta(\vx') \rangle \, e^{i\vk.(\vx-\vx')} .
\label{eqdelksq1}
\end{equation}
Substituting in Equations \ref{eq2pt} and \ref{eqpk} to Equation
\ref{eqdelksq1} we find, in a slight generalization of the results of
\citet{FKP} to include a general noise term,
\begin{equation}
  \langle |\tilde{\delta}(\vk)|^2 \rangle = \int
  \frac{d^3\vk'}{(2\pi)^3} \, P(\vk') \, |\tilde{W}(\vk-\vk')|^2 +
  \frac{1}{V} \int d^3\vx \, w^2(\vx) \, \sigma^2(\vx) = P \star
  |\tilde{W}|^2 + \overline{S} ,
\label{eqdelksq2}
\end{equation}
where we have defined a window function $W(\vx) = w(\vx) \, \langle
f(\vx) \rangle$.  Hence, $\langle |\tilde{\delta}(\vk)|^2 \rangle$ is
the sum of the convolution (which we denote by $\star$) of the
underlying power spectrum $P(\vk)$ and $|\tilde{W}(\vk)|^2$, and a
noise term,
\begin{equation}
  \overline{S} = \frac{1}{V} \int d^3\vx \, w^2(\vx) \, \sigma^2(\vx) .
\label{eqsdef}
\end{equation}
Repeating this process for the Fourier transform of two different
fluctuation fields \citep[see also][]{Smith09,Blake13}, weighted by
functions $w_1(\vx)$ and $w_2(\vx)$, we find that,
\begin{equation}
  \langle \tilde{\delta}_1(\vk) \, \tilde{\delta}_2^*(\vk) \rangle =
  \int \frac{d^3\vk'}{(2\pi)^3} \, P_c(\vk') \, \tilde{W}_1(\vk-\vk')
  \, \tilde{W}_2^*(\vk-\vk') = P_c \star \tilde{W}_1 \, \tilde{W}_2^*
  ,
\label{eqdelkcross}
\end{equation}
where $W_i(\vx) = w_i(\vx) \, \langle f_i(\vx) \rangle$.  In an
approximation where the power spectrum does not vary significantly
over the width of $|\tilde{W}(\vk)|^2$, such that we can take it
outside the integral over $\vk'$ in Equations \ref{eqdelksq2} and
\ref{eqdelkcross}, and using Parseval's theorem $\frac{V}{(2\pi)^3}
\int d^3\vk \, |\tilde{W}(\vk)|^2 = \frac{1}{V} \int d^3\vx \,
W^2(\vx)$, we find,
\begin{equation}
\langle |\tilde{\delta}(\vk)|^2 \rangle \approx \overline{Q} \,
\frac{P(\vk)}{V} + \overline{S} , \hspace{1cm} \langle
\tilde{\delta}_1(\vk) \, \tilde{\delta}_2^*(\vk) \rangle \approx
\overline{Q}_c \, \frac{P_c(\vk)}{V} ,
\label{eqdelkapprox}
\end{equation}
where we have defined dimensionless quantities,
\begin{equation}
\overline{Q} = \frac{1}{V} \int d^3\vx \, W^2(\vx) , \hspace{1cm}
\overline{Q}_c = \frac{1}{V} \int d^3\vx \, W_1(\vx) \, W_2(\vx) .
\label{eqqdef}
\end{equation}
In the following subsections we consider two important special cases
of fluctuation fields, which will be relevant in the subsequent
analysis.

\subsubsection{Poisson point process}

If $f(\vx)$ is generated by a Poisson point process from a galaxy
number density distribution $n_g(\vx)$, then $\xi(\vs) = 0$ and
\begin{equation}
\langle \delta(\vx) \, \delta(\vx') \rangle = V \, \langle n_g(\vx)
\rangle \, \delta_D(\vx-\vx') ,
\label{eqpoiss}
\end{equation}
and from the definition $\delta(\vx) = V \left[ n_g(\vx) - \langle n_g(\vx)
  \rangle \right]$,
\begin{equation}
\langle \delta(\vx) \, \delta(\vx') \rangle = V^2 \left[ \langle
  n_g(\vx) \, n_g(\vx') \rangle - \langle n_g(\vx) \rangle \, \langle
  n_g(\vx') \rangle \right] .
\end{equation}
To justify Equation \ref{eqpoiss}, we can use $N = \int d^3\vx \,
n_g(\vx)$ and consider
\begin{equation}
\langle N^2 \rangle = \int d^3\vx \int d^3\vx' \langle n_g(\vx)
n_g(\vx') \rangle = \int d^3\vx \int d^3\vx' \left[ \langle n_g(\vx)
  \rangle \langle n_g(\vx') \rangle + \frac{\langle n_g(\vx)
    \rangle}{V} \delta_D(\vx-\vx') \right] = \langle N \rangle^2 +
\int d^3\vx \langle n_g(\vx) \rangle = \langle N \rangle^2 + \langle N
\rangle ,
\end{equation}
as expected from Poisson statistics.  Comparing Equations \ref{eq2pt}
and \ref{eqpoiss}, we identify $\sigma^2(\vx) = V \, \langle n_g(\vx)
\rangle$ and hence assuming weight $w = 1$,
\begin{equation}
\langle |\tilde{\delta}(\vk)|^2 \rangle = \overline{S} = \int d^3\vx
\, \langle n_g(\vx) \rangle = N .
\end{equation}

\subsubsection{Uniform window and noise}
\label{secuniform}

Suppose that a field is sampled from a constant mean $\langle f(\vx)
\rangle = f_0$ with a constant noise $\sigma^2(\vx) = \sigma_0^2$,
where the weight function $w(\vx)$ takes the value $0$ (outside the
footprint) or 1 (inside the footprint), such that $w^2(\vx) = w(\vx)$
and the observed volume is $V_w = \int d^3\vx \, w(\vx)$.  In this
case, we find from Equations \ref{eqsdef} and \ref{eqqdef} that
$\overline{S} = \sigma_0^2 V_w / V$ and $\overline{Q} = f_0^2 V_w / V$
such that Equation \ref{eqdelkapprox} takes the form,
\begin{equation}
\langle |\tilde{\delta}(\vk)|^2 \rangle = \frac{f_0^2 V_w}{V^2} \left[
  P(\vk) + \frac{V \sigma_0^2}{f_0^2} \right] .
\end{equation}
In this scenario, the equivalent power spectrum due to noise is hence
the second term in the bracket,
\begin{equation}
  P_{\rm noise}(\vk) = \frac{V \sigma_0^2}{f_0^2} .
\label{eqpknoise}
\end{equation}
We note that for Poisson statistics, $f_0 = V n_0$ in terms of the
number density $n_0$, and $\sigma_0^2 = V n_0$, such that $P_{\rm
  noise} = 1/n_0$, as expected.

\subsection{Smoothing}
\label{secsmooth}

We now extend the power spectrum model to describe the effect of a
general smoothing of the fields, such as might result from a telescope
beam in radio observations, redshift errors in optical observations,
or a general pixelization.  We suppose that the smoothed fluctuation
field $\delta^{sm}(\vx)$ may be written in the form,
\begin{equation}
\delta^{sm}(\vx) = \frac{1}{V} \int d^3\vx' \, \delta(\vx') \,
B(\vx-\vx',\vx) ,
\label{eqsmoothing}
\end{equation}
where the dimensionless smoothing function $B$ is a compact function
of the separation $\vx-\vx'$, which may also vary with position $\vx$.
The smoothing function is normalized such that $\frac{1}{V} \int
d^3\vs \, B(\vs,\vx) = 1$ for all $\vx$, and we define the Fourier
transform of the smoothing function at each location as
$\tilde{B}(\vk,\vx) = \frac{1}{V} \int d^3\vs \, B(\vs,\vx) \,
e^{i\vk.\vs}$, where $\tilde{B}(\vk=0,\vx) = 1$.  Substituting in
these expressions we find that the Fourier transform of Equation
\ref{eqsmoothing} is,
\begin{equation}
  \tilde{\delta}^{sm}(\vk) = \int \frac{d^3\vk'}{(2\pi)^3} \,
  \tilde{\delta}(\vk') \int d^3\vx \, \tilde{B}(\vk',\vx) \,
  e^{i(\vk-\vk').\vx} ,
\end{equation}
which reproduces the standard result of the convolution theorem, that
$\tilde{\delta}^{sm}(\vk) = \tilde{B}(\vk) \, \tilde{\delta}(\vk)$ if
$B(\vs,\vx)$ is independent of position $\vx$.  The power spectrum of
the smoothed field is then,
\begin{equation}
\langle |\tilde{\delta}^{sm}(\vk)|^2 \rangle = \int
\frac{d^3\vk'}{(2\pi)^3} \langle |\tilde{\delta}(\vk')|^2 \rangle \int
d^3\vs \, e^{i(\vk-\vk').\vs} \, \frac{1}{V} \int d^3\vx \,
\tilde{B}(\vk',\vx) \, \tilde{B}^*(\vk',\vx+\vs) .
\end{equation}
Assuming that the smoothing function varies more slowly than the
clustering scale, we may utilize the approximation $\int d^3\vx \,
\tilde{B}(\vk,\vx) \, \tilde{B}^*(\vk,\vx+\vs) \approx \int d^3\vx \,
|\tilde{B}(\vk,\vx)|^2$.  Following this approximation the integral
over $\vs$ produces a delta function in $\vk-\vk'$, which leads to the
result that,
\begin{equation}
  \langle |\tilde{\delta}^{sm}(\vk)|^2 \rangle \approx \langle
  |\tilde{\delta}(\vk)|^2 \rangle \, \frac{1}{V} \int d^3\vx \,
  |\tilde{B}(\vk,\vx)|^2 = \langle |\tilde{\delta}(\vk)|^2 \rangle \,
  D^2(\vk) ,
\label{eqdamping}
\end{equation}
such that the power spectra (including both the signal and noise) are
modulated by a damping function $D^2(\vk) = \frac{1}{V} \int d^3\vx \,
|\tilde{B}(\vk,\vx)|^2$, which is the volume average of
$|\tilde{B}(\vk,\vx)|^2$.  If a smoothed field $\delta_1^{sm}(\vx)$ is
correlated with an unsmoothed field $\delta_2(\vx)$, the modulation of
the resulting cross-power spectrum is,
\begin{equation}
  \langle \tilde{\delta}_1^{sm}(\vk) \, \tilde{\delta}_2^*(\vk)
  \rangle \approx \langle \tilde{\delta}_1(\vk) \,
  \tilde{\delta}_2^*(\vk) \rangle \, \frac{1}{V} \int d^3\vx \,
  \tilde{B}(\vk,\vx) ,
\label{eqdampingcross}
\end{equation}
or for the cross-correlation of two fields smoothed with different
functions $B_1(\vs,\vx)$ and $B_2(\vs,\vx)$,
\begin{equation}
  \langle \tilde{\delta}_1^{sm}(\vk) \, \tilde{\delta}_2^{sm \,
    *}(\vk) \rangle \approx \langle \tilde{\delta}_1(\vk) \,
  \tilde{\delta}_2^*(\vk) \rangle \, \frac{1}{V} \int d^3\vx \,
  \tilde{B}_1(\vk,\vx) \, \tilde{B}_2^*(\vk,\vx) .
\end{equation}
In the following subsections we consider some special cases of these
results, which will be utilized in the subsequent N-body simulation
tests.

\subsubsection{Pixelization}
\label{secpix}

A special case of the smoothing operation described by Equation
\ref{eqsmoothing} occurs when a field is pixelized into distinct
``cells'', such that the average value of the field within each cell
$i$ is assigned to all positions within the cell.  This behaviour can
be modelled if the field is averaged by a top-hat function centred on
each pixel position $\vx_i$ such that,
\begin{equation}
\delta^{sm}(\vx) = \sum_i \frac{1}{\Delta V_i} \, P_i(\vx) \int_{{\rm
    cell} \, i} d^3\vx' \, \delta(\vx') = \frac{1}{V} \int_{\rm all}
d^3\vx' \delta(\vx') \sum_i \frac{V}{\Delta V_i} \, P_i(\vx) \,
P_i(\vx') ,
\end{equation}
where $\Delta V_i$ is the volume of the cell and $P_i(\vx) = 1$ if
$\vx$ is in cell $i$, and zero otherwise.  We now define the cell
window function $T_i(\vx - \vx_i) = 1$ if $\vx$ is in cell $i$, and
zero otherwise, and an offset function with respect to the pixel
position, $\ve_i(\vx) = \vx - \vx_i$.  Hence we can identify by
comparison with Equation \ref{eqsmoothing},
\begin{equation}
B(\vx-\vx',\vx) = \sum_i \frac{V}{\Delta V_i} \, P_i(\vx) \,
T_i(\vx'-\vx+\ve_i(\vx)) .
\end{equation}
Taking the Fourier transform of this equation,
\begin{equation}
\tilde{B}(\vk,\vx) = \sum_i \frac{1}{\Delta V_i} \, P_i(\vx) \int
d^3\vs \, T_i(\vs + \ve_i(\vx)) \, e^{i\vk.\vs} = \sum_i
\frac{V}{\Delta V_i} \, P_i(\vx) \, \tilde{T}_i(\vk) \, e^{-i
  \vk.\ve_i(\vx)} .
\end{equation}
In order to interpret this equation, we note that $\tilde{T}_i(0) =
\Delta V_i/V$ and $\tilde{T}_i(\vk)$ is proportional to
$\tilde{B}(\vk,\vx)$ at the position of cell $i$.  Hence, pixelization
results in a phase change in the Fourier transform such that,
\begin{equation}
 \tilde{B}(\vk,\vx) \rightarrow \tilde{B}(\vk,\vx) \,
 e^{-i\vk.\ve(\vx)} .
\end{equation}
This behaviour does not change the value of $|\tilde{B}(\vk,\vx)|^2$,
hence the overall effect on the auto-power spectrum can still be
evaluated using Equation \ref{eqdamping}.  However, the damping of the
cross-power spectrum as computed by Equation \ref{eqdampingcross} is
changed by this type of smoothing, by the volume average of
$e^{-i\vk.\ve(\vx)}$.  Given that the offsets $\ve(\vx)$ will be
distributed across space with the same profile as the cells, this
volume average is well-approximated by $\tilde{B}^*(\vk,\vx)$.
Therefore, even though only one of the two fields is smoothed, the
cross-power spectrum is damped due to pixelization by approximately
the same factor $D^2(\vk)$ as the auto-power spectrum.

\subsubsection{Noise applied to cells}

We now consider a scenario where the noise in the fluctuation field
is generated by drawing a random variable in a series of cells $i$ of
volume $\Delta V_i$, with zero mean and variance $\sigma^2_i$.  In
this case, the 2-point statistics of the noise in Equation \ref{eq2pt}
is modified from $\langle \delta(\vx) \, \delta(\vx') \rangle =
\sigma^2(\vx) \, \delta_D(\vx-\vx')$ to,
\begin{equation}
\langle \delta(\vx) \, \delta(\vx') \rangle = \sum_i \sigma_i^2 \,
P_i(\vx) \, P_i(\vx') ,
\end{equation}
such that the noise is uncorrelated between different cells.
Substituting this relation in Equation \ref{eqdelksq1} we find,
\begin{equation}
\langle |\tilde{\delta}(\vk)|^2 \rangle = \frac{1}{V^2} \sum_i
\sigma_i^2 \int d^3\vx \int d^3\vx' \, w(\vx) \, w(\vx') \, P_i(\vx)
\, P_i(\vx') \, e^{i\vk.(\vx-\vx')} = \frac{1}{V^2} \sum_i \sigma_i^2
\, w_i^2 \, |\tilde{B}(\vk,\vx_i)|^2 \, (\Delta V_i)^2 ,
\label{eqdelksqnoise}
\end{equation}
after applying the same arguments as in the previous subsection, where
$w_i$ is the weight applied to cell $i$.  The noise power spectrum
(Equation \ref{eqsdef}) including pixelization,
\begin{equation}
\langle |\tilde{\delta}(\vk)|^2 \rangle = \frac{1}{V} \int d^3\vx \,
w^2(\vx) \, \sigma^2(\vx) \, |\tilde{B}(\vk,\vx)|^2 ,
\end{equation}
can then be recovered by comparison with Equation \ref{eqdelksqnoise}
if we define the appropriate noise variance,
\begin{equation}
  \sigma^2(\vx) = \frac{1}{V} \sum_i \sigma_i^2 \, \Delta V_i \,
  P_i(\vx) .
\label{eqnoisepix}
\end{equation}
Poisson noise, $\sigma^2(\vx) = V \langle n_g(\vx) \rangle$, is
obtained if $\sigma_i^2 = V^2 \langle n_{g,i} \rangle / \Delta V_i$,
where $\langle n_{g,i} \rangle$ is the galaxy number density in cell
$i$.  This result will be useful in Section \ref{secsim}, for adding a
Poisson noise component to model an intensity map constructed by
binning a simulation catalogue of discrete objects in cells.

\subsubsection{Telescope beam}

For Gaussian smoothing perpendicular to the line-of-sight, such as
would result from a radio telescope beam, we have a smoothing kernel
$B(s_\perp) \propto e^{-s_\perp^2/2\sigma_\perp^2}$ as a function of
perpendicular spatial separation $s_\perp$, where $\sigma_\perp$ is
the spatial standard deviation of the beam.  The Fourier transform of
this function is,
\begin{equation}
  \tilde{B}_{\rm beam}(\vk) = e^{-k_\perp^2 \sigma_\perp^2/2} ,
\label{eqbbeam}
\end{equation}
where $k_\perp = k \sqrt{1 - \mu^2}$ and $\mu$ is the cosine of the
angle between $\vk$ and the line-of-sight.  Hence, the beam damps
power at small perpendicular separations.  For a beam of constant
angular standard deviation $\sigma_\theta$ on the sky in units of
radians, the corresponding spatial smoothing scale will vary with
position as $\sigma_\perp(\vx) = |\vx| \, \sigma_\theta$.  In this
case we would derive the damping factor using Equation \ref{eqdamping}
as,
\begin{equation}
D^2(\vk) = \frac{1}{V} \int d^3\vx \, e^{-k_\perp^2 \, |\vx|^2 \,
  \sigma_\theta^2} .
\end{equation}

\subsubsection{Frequency channels}

Smoothing in the radial direction results from the width of the
frequency channels in which radio intensity mapping data is collected.
For a frequency channel of spatial width $s_\parallel$ the Fourier
transform of the top-hat assignment function is,
\begin{equation}
  \tilde{B}_{\rm chan}(\vk) = \frac{\sin{(k_\parallel
      s_\parallel/2)}}{k_\parallel s_\parallel/2} ,
\end{equation}
where $k_\parallel = k \mu$.  If the frequency width of the channel is
$\Delta \nu$ then, for line measurements with rest frequency $\nu_0$
such that $\nu = \nu_0/(1+z)$, $s_\parallel(\vx) = [c/H(z)] \, (1+z)^2
\, (\Delta \nu/\nu_0)$ in terms of the speed of light $c$ and Hubble
parameter $H(z)$.  In this case we would derive the damping factor
using Equation \ref{eqdamping} as,
\begin{equation}
D^2(\vk) = \frac{1}{V} \int d^3\vx \, \left[ \frac{\sin{(k_\parallel
      s_\parallel(\vx)/2)}}{k_\parallel s_\parallel(\vx)/2} \right]^2
.
\end{equation}

\subsubsection{Redshift errors}

Damping of power in the radial direction can also result from errors
in measured galaxy redshifts, for example due to photometric redshift
estimates with error $\Delta z$ (which may vary with redshift).
Assuming that these errors are Gaussian, the Fourier transform of the
smoothing kernel follows Equation \ref{eqbbeam},
\begin{equation}
  \tilde{B}_{\Delta z}(\vk) = e^{-k_\parallel^2 \sigma_\parallel^2/2} ,
\end{equation}
where $\sigma_\parallel(\vx) = [c/H(z)] \, \Delta z$ is the spatial
radial error in each location.  In this case the overall damping
factor can be computed using,
\begin{equation}
D^2(\vk) = \frac{1}{V} \int d^3\vx \, e^{-k_\parallel^2 \,
  \sigma_\parallel^2(\vx)} .
\end{equation}
\citet{Chaves18} present a related investigation of the effect of
photo-$z$ errors on multipoles of the baryon acoustic oscillation
power spectrum.

\subsubsection{Angular pixelization}

A process of angular pixelization, for example using a scheme such as
{\tt HEALPix} \citep{Healpix05}, results in a damping of power as a
function of $k_\perp$.  The associated damping of the angular power
spectrum $C_\ell$ of the field as a function of multipole $\ell$ can
be expressed in terms of the pixel window function $W_{\rm
  ang}(\ell)$, such that the damping is described by $C_\ell
\rightarrow C_\ell \, W_{\rm ang}^2(\ell)$ and
\begin{equation}
W_{\rm ang}^2(\ell) = \frac{4\pi}{2\ell+1} \sum_{m=-\ell}^\ell
|w_{\ell m}|^2 ,
\end{equation}
where $w_{\ell m} = \int_{\rm pixel} d\Omega \, Y_{\ell m}(\Omega)$ is
the spherical harmonic transform of a pixel in terms of the spherical
harmonic functions $Y_{\ell m}$.  In the case of a 3D survey smoothed
using angular pixelization, the contribution to the damping factor at
position $\vx$ relative to the observer is determined by identifying
$\ell = k_\perp |\vx|$ such that $\tilde{B}_{\rm ang}(\vk,\vx) =
W_{\rm ang}(k_\perp |\vx|)$.  In this case we would derive the damping
factor for the 3D power spectrum as,
\begin{equation}
D^2(\vk) = \frac{1}{V} \int d^3\vx \, W_{\rm ang}^2(k_\perp |\vx|) .
\end{equation}

\subsubsection{Nearest grid point assignment}

For nearest grid point assignment, the Fourier transform of the
assignment function is,
\begin{equation}
  \tilde{B}_{\rm NGP}(\vk) = \frac{\sin{(k_xH/2)}}{(k_xH/2)} \,
  \frac{\sin{(k_yH/2)}}{(k_yH/2)} \, \frac{\sin{(k_zH/2)}}{(k_zH/2)} ,
\label{eqngpsmooth}
\end{equation}
where $H$ is the grid spacing \citep[e.g.,][]{Jing05}.  This special
case will be useful in the following section.

\subsection{Discretization}
\label{secdisc}

We now consider the effect on the power spectrum if a continuous field
$\delta(\vx)$, possibly having been smoothed using one of the schemes
described in the previous section, is sampled on a regular FFT grid at
positions $\vx_\vn = H \vn$, where $\vn$ is a vector of integers.
This case is important for efficient power spectrum estimators.
Following \citet{Jing05}, we can conveniently describe this process
using the sampling function $\Pi(\vx) = \sum_\vn \delta_D(\vx-\vn)$,
an array of $\delta$-functions placed at integers $\vn$, such that the
gridded field $\delta^{gr}(\vx)$ can be written in the form,
\begin{equation}
  \delta^{gr}(\vx) = \Pi(\vx/H) \, \delta(\vx) .
\label{eqsamp}
\end{equation}
We can see that Equation \ref{eqsamp} produces the correct result for
the Fourier-transformed field by considering,
\begin{equation}
\tilde{\delta}^{gr}(\vk) = \frac{1}{V} \int d^3\vx \, \delta^{gr}(\vx)
\, e^{i\vk.\vx} = \sum_\vn \frac{1}{V} \int d^3\vx \, \delta_D(\vx -
\vn H) \, \delta(\vx) \, e^{i\vk.\vx} = \sum_\vn \delta(\vx_\vn) \,
e^{i\vk.\vx_\vn} ,
\end{equation}
as expected when evaluating an FFT.  The Fourier transform of Equation
\ref{eqsamp} may also be obtained using the convolution theorem,
\begin{equation}
\tilde{\delta}^{gr}(\vk) = \frac{V}{(2\pi)^3} \int d^3\vk' \,
\tilde{\delta}(\vk') \, \tilde{\Pi}(\vk-\vk') ,
\label{eqsampconv}
\end{equation}
where the Fourier transform of the sampling function is given by,
\begin{equation}
\tilde{\Pi}(\vk) = \frac{1}{V} \int d^3\vx \, \Pi(\vx/H) \,
e^{i\vk.\vx} = \sum_\vn e^{i\vk.\vn H} = \sum_\vn e^{i 2\pi
  \frac{\vk.\vn}{2k_N}} = \sum_\vn \tilde{\delta}_D(\vk - 2k_N \vn) ,
\end{equation}
where $k_N = \pi/H$ is the Nyquist frequency of the grid.  Hence,
Equation \ref{eqsampconv} may be simplified as,
\begin{equation}
\tilde{\delta}^{gr}(\vk) = \sum_{\vk_\vn} \tilde{\delta}(\vk_\vn) ,
\end{equation}
where $\vk_\vn = \vk + 2k_N\vn$ such that,
\begin{equation}
\langle |\tilde{\delta}^{gr}(\vk)|^2 \rangle = \sum_{\vk_\vn} \langle
|\tilde{\delta}(\vk_\vn)|^2 \rangle ,
\label{eqdisc}
\end{equation}
where cross-terms disappear because of homogeneity.  Hence,
discretization involves the aliasing of power to scale $\vk$ from a
series of scales $\vk_\vn$ spaced by $2 k_N$ \citep{Jing05}.
Discretization is often combined with smoothing, in which case
combining Equations \ref{eqdamping} and \ref{eqdisc} yields,
\begin{equation}
\langle |\tilde{\delta}^{gr}(\vk)|^2 \rangle = \sum_{\vk_\vn} \langle
|\tilde{\delta}(\vk_\vn)|^2 \rangle \, D^2(\vk_\vn) .
\end{equation}
We note an interesting special case in which nearest grid point
assignment is applied to a Poisson noise spectrum $\langle
|\tilde{\delta}(\vk)|^2 \rangle = \overline{S}$, in which case the
power spectrum is unchanged:
\begin{equation}
\langle |\tilde{\delta}^{gr}(\vk)|^2 \rangle = \overline{S} \,
\sum_{\vk_\vn} D_{\rm NGP}^2(\vk_\vn) = \overline{S} ,
\end{equation}
where we have used the identity $\sum_{\vk_\vn} |\tilde{B}_{\rm
  NGP}(\vk_\vn)|^2 = 1$, where $\tilde{B}_{\rm NGP}(\vk)$ is defined
by Equation \ref{eqngpsmooth}.\footnote{This identity is known as
  Glaisher's series.}

Combining the results of the above sections, we can describe the joint
effects of the window function, noise, smoothing and discretization on
the auto-power spectrum of the fluctuation field by,
\begin{equation}
\langle |\tilde{\delta}(\vk)|^2 \rangle = \sum_{\vk_\vn} \left[ (P
  \star |\tilde{W}|^2)(\vk_\vn) + \overline{S} \right] D^2(\vk_\vn) ,
\end{equation}
and on the cross-power spectrum by,
\begin{equation}
  \langle \tilde{\delta}_1(\vk) \, \tilde{\delta}_2^*(\vk) \rangle =
  \sum_{\vk_\vn} \left[ (P_c \star \tilde{W}_1
    \tilde{W}_2^*)(\vk_\vn) \right] D^2(\vk_\vn) .
\end{equation}

\subsection{Power spectrum multipoles}
\label{secmult}

As the above sections demonstrate, the contribution of a fluctuation
field to its power spectrum in a volume may vary as a function of
$\vx$ owing to variations in the statistical properties of the field
such as its clustering, noise, smoothing or weighting.  We can
encapsulate these effects by writing the observed power spectrum as an
integral over $\vx$,
\begin{equation}
  P(\vk) = \frac{1}{V} \int d^3\vx \, P(\vk,\vx) ,
\label{eqpkx}
\end{equation}
where $P(\vk,\vx)$ represents the contribution to the power spectrum
originating from position $\vx$.

This concept is useful when considering a further cause of
position-dependence: the changing line-of-sight direction across the
volume.  Effects such as redshift-space distortions, the telescope
beam and angular/radial smoothing will cause the amplitude of the
power spectrum $P(\vk)$ to depend on the direction of $\vk$ with
respect to a global line-of-sight.  Assuming azimuthal symmetry, the
power spectrum will only depend on $\mu$, the cosine of the angle
between $\vk$ and the line-of-sight direction.  In this case the 2D
function $P(k,\mu)$ may be conveniently quantified by power spectrum
multipoles, $P_\ell(k)$:
\begin{equation}
P(k,\mu) = \sum_\ell P_\ell(k) \, L_\ell(\mu) = \sum_\ell P_\ell(k) \,
L_\ell(\hk.\hx) ,
\label{eqpkmult}
\end{equation}
where $L_\ell$ are the Legendre polynomials, and in the last
expression we are describing the varying line-of-sight, given that in
a region of space around position $\vx$ from the observer we can write
$\mu = \hk.\hx$.  Inverting Equation \ref{eqpkmult} and averaging the
statistic over all positions using Equation \ref{eqpkx}, we can model
the power spectrum multipoles as,
\begin{equation}
  \begin{split}
    P_\ell(k) &= \frac{2\ell+1}{V} \int d^3\vx \int
    \frac{d\Omega_k}{4\pi} \, P(\vk,\vx) \, L_\ell(\hk.\hx) \\ &=
    (2\ell+1) \int \frac{d\Omega_k}{4\pi} \, \frac{1}{V} \int d^3\vx
    \int d^3\vx' \, \xi(\vx-\vx') \, e^{i\vk.(\vx-\vx')} \,
    L_\ell(\hk.\hx') ,
  \end{split}
\end{equation}
where $d\Omega_k$ integrates over all angles $\hk$.  The equivalent
relation to Equation \ref{eqdelksq1} for a power spectrum multipole
$\ell$ can then be written as,
\begin{equation}
\langle |\tilde{\delta}(\vk)|^2_\ell \rangle = \frac{1}{V^2} \int
d^3\vx \int d^3\vx' \, w(\vx) \, w(\vx') \, \langle \delta(\vx) \,
\delta(\vx') \rangle \, e^{i\vk.(\vx-\vx')} \, L_\ell(\hk.\hx') ,
\end{equation}
where $\langle |\tilde{\delta}(\vk)|^2_\ell \rangle$ describes the
contribution of wavenumber $\vk$ to the power spectrum multipole
$P_\ell(k)$.  \citet{Blake18} develop the equivalent expressions to
Equations \ref{eqdelksq2} and \ref{eqdelkcross} for modelling the
auto- and cross-power spectrum multipoles.  For the auto-power
spectrum:
\begin{equation}
\langle |\tilde{\delta}(\vk)|^2_\ell \rangle = \sum_{\ell'} \int
\frac{d^3\vk'}{(2\pi)^3} \, P_{\ell'}(k') \, \tilde{W}(\vk-\vk') \,
W_{\ell \ell'}^*(\vk,\vk') + \frac{1}{V} \int d^3\vx \, w^2(\vx) \,
\sigma^2(\vx) \, L_\ell(\hk.\hx) ,
\end{equation}
where
\begin{equation}
W_{\ell \ell'}(\vk,\vk') = \frac{1}{V} \int d^3\vx \, W(\vx) \,
L_\ell(\hk.\hx) \, L_{\ell'}(\hk'.\hx) \, e^{i(\vk-\vk').\vx} .
\end{equation}
The equivalent expression for the cross-power spectrum multipoles is,
\begin{equation}
\langle \tilde{\delta}_1(\vk) \, \tilde{\delta}_2^*(\vk)_\ell \rangle
= \sum_{\ell'} \int \frac{d^3\vk'}{(2\pi)^3} \, P_{c,\ell'}(k') \,
\tilde{W}_1(\vk-\vk') \, W_{2, \ell \ell'}^*(\vk,\vk') .
\end{equation}
These equations reduce to the results of Equations \ref{eqdelksq2} and
\ref{eqdelkcross} for $\ell = \ell' = 0$.

\section{Power spectrum measurement}
\label{secmeas}

In this section we consider the estimators for the auto- and
cross-power spectra, the variance in these estimators, and the optimal
weighting of the fluctuation fields which minimizes this variance
under certain approximations.  These results are useful for practical
power spectrum analysis.

\subsection{Estimators}
\label{secpkest}

Equation \ref{eqdelkapprox} motivates an estimator for the power
spectrum in terms of the Fourier transform of the weighted fluctuation
field, $\tilde{\delta}(\vk)$,
\begin{equation}
  \hP(\vk) = \frac{\left( |\tilde{\delta}(\vk)|^2 - \overline{S}
    \right) V}{\overline{Q}} ,
  \label{eqpkautoest}
\end{equation}
such that $\langle \hP(\vk) \rangle \approx P(\vk)$.  Similarly for
the cross-power spectrum,
\begin{equation}
\hP_c(\vk) = \frac{{\rm Re} \lbrace \tilde{\delta}_1(\vk) \,
  \tilde{\delta}_2^*(\vk) \rbrace \, V}{\overline{Q}_c} ,
  \label{eqpkcrossest}
\end{equation}
where $\langle \hP_c(\vk) \rangle \approx P_c(\vk)$ and ${\rm Re}
\lbrace \rbrace$ indicates we are taking the real part of the
expression, noting that Equation \ref{eqpkcrossest} is symmetric in
the two fields since ${\rm Re} \lbrace \tilde{\delta}_1 \,
\tilde{\delta}_2^* \rbrace = \left( \tilde{\delta}_1 \,
\tilde{\delta}_2^* + \tilde{\delta}_1^* \, \tilde{\delta}_2
\right)/2$.  The estimator for the power spectrum multipoles is,
\begin{equation}
\hP_\ell(\vk) = \frac{(2\ell+1) \left( |\tilde{\delta}(\vk)|^2_\ell -
  S_\ell(\vk) \right) V}{\overline{Q}} ,
  \label{eqpkmultest}
\end{equation}
where $S_\ell(\vk) = \frac{1}{V} \int d^3\vx \, w^2(\vx) \,
\sigma^2(\vx) \, L_\ell(\hk.\hx)$.  \citet{Bianchi15} provide an
FFT-based method for evaluating Equation \ref{eqpkmultest} \citep[see
  also,][]{Scoccimarro15}.  Examples of clustering analyses where
these estimators are used include \citet{GilMarin16},
\citet{Beutler17}, \citet{GilMarin18} and \citet{Blake18}.  The
estimator for the cross-power spectrum multipoles is,
\begin{equation}
\hP_{c,\ell}(\vk) = \frac{(2\ell+1) \, {\rm Re} \lbrace
  \tilde{\delta}_1(\vk) \, \tilde{\delta}_2^*(\vk)_\ell \rbrace \,
  V}{\overline{Q}_c}
\label{eqpkcrossmultest}
\end{equation}
which can be evaluated using an adapted version of the
\citet{Bianchi15} method.  Equations \ref{eqpkautoest} to
\ref{eqpkcrossmultest} all provide estimates of the corresponding
power spectrum for a mode with wavenumber $\vk$.  We can then bin
these estimates in spherical shells of $k = |\vk|$, to extract
measurements of mode-averaged (multipole) power spectra.

\subsection{Errors and optimal weighting}
\label{secpkerr}

We now consider the variance in these power spectrum estimators.
Assuming Gaussian statistics, the covariance of the power spectrum
estimate between two modes $\vk$ and $\vk'$ separated by $\delta \vk =
\vk - \vk'$ is given by \citep[see][]{FKP,Blake13},
\begin{equation}
\langle \delta \hP(\vk) \, \delta \hP(\vk') \rangle \approx
\frac{|P(\vk) \, \tilde{Q}(\delta \vk) + V \, \tilde{S}(\delta
  \vk)|^2}{\overline{Q}^2} ,
\label{eqpkcov}
\end{equation}
where $\delta \hP(\vk)$ is the fluctuation in value of the power
spectrum estimator, and $\tilde{Q}(\vk)$ and $\tilde{S}(\vk)$ are the
Fourier transforms of $Q(\vx) = w^2(\vx) \, \langle f(\vx) \rangle^2$
and $S(\vx) = w^2(\vx) \, \sigma^2(\vx)$, respectively, such that
$\overline{Q} \equiv \tilde{Q}(0)$ and $\overline{S} \equiv
\tilde{S}(0)$.  If we average the power spectrum estimates in a bin of
Fourier space of volume $V_k$ near wavenumber $\vk$, this produces a
variance in a bin which may be approximately evaluated as \citep{FKP},
\begin{equation}
\sigma_P^2 \approx \frac{1}{V_k} \int d^3\vk' \, \frac{|P(\vk) \,
  \tilde{Q}(\vk') + V \, \tilde{S}(\vk')|^2}{\overline{Q}^2} ,
\end{equation}
assuming the width of the bin is large compared to the correlation
length in $\vk$-space.  Using Parseval's theorem, together with the
expression for the number of unique modes in the bin $N_m = V_k V / (2
\pi)^3$, this yields,
\begin{equation}
\sigma_P^2 \approx \frac{1}{N_m \, V} \int d^3\vx \, \frac{\left[
    P(\vk) \, Q(\vx) + V \, S(\vx) \right]^2}{\overline{Q}^2} =
\frac{V^3}{N_m} \frac{\int d^3\vx \, w^4(\vx) \left[ \frac{P(\vk)}{V}
    \, \langle f(\vx) \rangle^2 + \sigma^2(\vx) \right]^2}{\left[ \int
    d^3\vx \, w^2(\vx) \, \langle f(\vx) \rangle^2 \right]^2} .
\label{eqsigp}
\end{equation}
In the special case corresponding to Section \ref{secuniform}, where
the field is sampled from a constant mean $\langle f(\vx) \rangle =
f_0$ with a constant noise $\sigma^2(\vx) = \sigma_0^2$, within a
subset of the cuboid defined by $w = 1$ we find,
\begin{equation}
\sigma_P^2 = \frac{1}{N_m} \, \frac{V}{V_w} \, \left[ P(\vk) + \frac{V
    \sigma_0^2}{f_0^2} \right]^2 ,
\label{eqpkautoerr}
\end{equation}
noting that for Poisson statistics, the second term in the bracket in
Equation \ref{eqpkautoerr} reduces to $1/n_0$.

Following \citet{FKP}, we can determine the weight function $w(\vx)$
in Equation \ref{eqsigp} which minimizes $\sigma_P^2$ by solving the
equation $\partial \sigma_P^2/\partial w = 0$.  We find:
\begin{equation}
  \frac{\partial \sigma_P^2}{\partial w} \propto \left[ \int d^3\vx \,
    4w^3 \left( \frac{P}{V} \langle f \rangle^2 + \sigma^2 \right)^2
    \right] \left[ \int d^3\vx \, w^2 \langle f \rangle^2 \right]^{-2}
  - 2 \left[ \int d^3\vx \, w^4 \left( \frac{P}{V} \langle f \rangle^2
    + \sigma^2 \right)^2 \right] \left[ \int d^3\vx \, w^2 \langle f
    \rangle^2 \right]^{-3} \left[ \int d^3\vx \, 2w \langle f
    \rangle^2 \right] = 0 ,
\end{equation}
which may be re-arranged to yield,
\begin{equation}
\frac{\int d^3\vx \, w^3 \, \left( \frac{P}{V} \, \langle f \rangle^2
  + \sigma^2 \right)^2}{\int d^3\vx \, w^4 \, \left( \frac{P}{V} \,
  \langle f \rangle^2 + \sigma^2 \right)^2} = \frac{\int d^3\vx \, w
  \, \langle f \rangle^2}{\int d^3\vx \, w^2 \, \langle f \rangle^2} .
\end{equation}
By inspection, we find that this equation is satisfied if,
\begin{equation}
  w^2 \, \left( \frac{P}{V} \, \langle f \rangle^2 + \sigma^2
  \right)^2 = \langle f \rangle^2 ,
\end{equation}
or,
\begin{equation}
  w(\vx) = \frac{\langle f(\vx) \rangle}{\frac{P(\vk)}{V} \, \langle
    f(\vx) \rangle^2 + \sigma^2(\vx)} .
\label{eqoptwei}
\end{equation}
In the case of a galaxy survey with Poisson statistics, we have
$\langle f(\vx) \rangle = \sigma^2(\vx) = V \langle n_g(\vx) \rangle$,
in which case we recover the usual dimensionless FKP weighting,
\begin{equation}
  w(\vx) = \frac{1}{1 + \langle n_g(\vx) \rangle \, P(\vk)} .
\label{eqoptweigal}
\end{equation}
For an intensity map with temperature variance $\sigma_T^2(\vx)$ and
mean brightness temperature $\langle f(\vx) \rangle = \langle T_b(\vx)
\rangle$ we have,
\begin{equation}
  w(\vx) = \frac{\langle T_b(\vx) \rangle}{\frac{P(\vk)}{V} \, \langle
    T_b(\vx) \rangle^2 + \sigma_T^2(\vx)} ,
\label{eqoptweitemp}
\end{equation}
which has dimensions of inverse temperature as required (see the
footnote after Equation \ref{eqdelk}).

The expression equivalent to Equation \ref{eqpkcov} for the
cross-power spectrum is \citep{Smith09,Blake13},
\begin{equation}
\langle \delta \hP_c(\vk) \, \delta \hP_c(\vk') \rangle \approx \frac{
  |P_c(\vk) \, \tilde{Q}_c(\delta \vk)|^2 + {\rm Re} \lbrace \left[
    P_1(\vk) \, \tilde{Q}_1(\delta \vk) + V \, \tilde{S}_1(\delta \vk)
    \right] \left[ P_2(\vk) \, \tilde{Q}_2(\delta \vk) + V \,
    \tilde{S}_2(\delta \vk) \right]^* \rbrace}{2 \, \overline{Q}_c^2}
,
\end{equation}
where $\tilde{Q}_c(\vk)$ is the Fourier transform of $Q_c(\vx) =
w_1(\vx) \, w_2(\vx) \, \langle f_1(\vx) \rangle \, \langle f_2(\vx)
\rangle$ and $\overline{Q}_c \equiv \tilde{Q}_c(0)$.  This leads to,
\begin{equation}
\begin{split}
  \sigma_{P_c}^2 &\approx \frac{1}{N_m \, V} \int d^3\vx \, \frac{
    \left( P_c^2(\vk) \, Q_c^2(\vx) + \left[ P_1(\vk) \, Q_1(\vx) + V
      \, S_1(\vx) \right] \, \left[ P_2(\vk) \, Q_2(\vx) + V \,
      S_2(\vx) \right] \right)}{2 \, \overline{Q}_c^2} \\ &=
  \frac{V^3}{2 \, N_m} \frac{\int d^3\vx \, w_1^2(\vx) \, w_2^2(\vx)
    \, \left( \frac{P_c^2(\vk)}{V^2} \, \langle f_1(\vx) \rangle^2 \,
    \langle f_2(\vx) \rangle^2 + \left[ \frac{P_1(\vk)}{V} \, \langle
      f_1(\vx) \rangle^2 + \sigma_1^2(\vx) \right] \, \left[
      \frac{P_2(\vk)}{V} \, \langle f_2(\vx) \rangle^2 +
      \sigma_2^2(\vx) \right] \right)}{\left[ \int d^3\vx \, w_1(\vx)
      \, w_2(\vx) \, \langle f_1(\vx) \rangle \, \langle f_2(\vx)
      \rangle \right]^2} .
\end{split}
\end{equation}
In the special case where the weights, means and noise are
position-independent we find,
\begin{equation}
\sigma_{P_c}^2 = \frac{1}{2 \, N_m} \, \frac{V}{V_{w,c}} \, \left[
  P_c^2 + \left( P_1 + \frac{V \sigma_1^2}{f_1^2} \right) \, \left(
  P_2 + \frac{V \sigma_2^2}{f_2^2} \right) \right] ,
\label{eqpkcrosserr}
\end{equation}
where $V_{w,c} = \int d^3\vx \, w_1(\vx) \, w_2(\vx)$ is the overlap
volume of the two datasets.  Following the same method as above, we
find that $\partial \sigma_{P_c}^2/\partial w_1 = \partial
\sigma_{P_c}^2/\partial w_2 = 0$, such that the error in the
cross-power spectrum is minimized, if the product of the individual
weights satisfies,
\begin{equation}
  w_1(\vx) \, w_2(\vx) = \frac{\langle f_1(\vx) \rangle \, \langle
    f_2(\vx) \rangle}{\frac{P_c^2(\vk)}{V^2} \, \langle f_1(\vx)
    \rangle^2 \, \langle f_2(\vx) \rangle^2 + \left[
      \frac{P_1(\vk)}{V} \, \langle f_1(\vx) \rangle^2 +
      \sigma_1^2(\vx) \right] \, \left[ \frac{P_2(\vk)}{V} \, \langle
      f_2(\vx) \rangle^2 + \sigma_2^2(\vx) \right]} .
\label{eqoptweicross}
\end{equation}
We note that assigning the weights for the two datasets according to
the optimal single-tracer weights (Equation \ref{eqoptwei}) does not
produce the optimal weight for the cross-power spectrum unless $P_c =
0$.  The optimal error in the cross-power spectrum, which also
satisfies Equation \ref{eqoptweicross}, can be produced if the
single-tracer optimal weights are modified such that $w_i'(\vx) =
w_c(\vx) \, w_i(\vx)$ where,
\begin{equation}
  w_c(\vx) = \left[ 1 + w_1(\vx) \, w_2(\vx) \, \frac{P_c^2(\vk)}{V^2}
    \, \langle f_1(\vx) \rangle \, \langle f_2(\vx) \rangle
    \right]^{-1/2} .
\end{equation}
In the case of galaxy surveys with Poisson statistics, these optimal
weights are
\begin{equation}
w_1(\vx) = \frac{1}{1+ \langle n_1(\vx) \rangle \, P_1(\vk)}
, \hspace{1cm} w_2(\vx) = \frac{1}{1+ \langle n_2(\vx) \rangle \,
  P_2(\vk)} , \hspace{1cm} w_c(\vx) = \frac{1}{\sqrt{1 + w_1(\vx) \,
    w_2(\vx) \, \langle n_1(\vx) \rangle \, \langle n_2(\vx) \rangle
    \, P_c^2(\vk)}} ,
\end{equation}
and in the case of the auto- and cross-correlations of a galaxy survey
and an intensity mapping survey,
\begin{equation}
w_g(\vx) = \frac{1}{1+ \langle n_g(\vx) \rangle \, P_g(\vk)}
, \hspace{1cm} w_T(\vx) = \frac{\langle T_b(\vx)
  \rangle}{\frac{P_T(\vk)}{V} \, \langle T_b(\vx) \rangle^2 +
  \sigma_T^2(\vx)} , \hspace{1cm} w_c(\vx) = \frac{1}{\sqrt{1 +
    w_g(\vx) \, w_T(\vx) \, \langle n_g(\vx) \rangle \, \langle
    T_b(\vx) \rangle \, \frac{P_c^2(\vk)}{V}}} .
\end{equation}

We obtain the error in the power spectrum multipoles by taking the
multipoles of Equations \ref{eqpkautoerr} and \ref{eqpkcrosserr}
\citep{Grieb16},
\begin{equation}
\begin{split}
  \sigma_{P_\ell}^2 &= (2\ell+1)^2 \, \frac{1}{N_m} \, \frac{V}{V_w}
  \, \int_0^1 d\mu \, \left[ P(k,\mu) + \frac{V \sigma_0^2}{f_0^2}
    \right]^2 L_\ell^2(\mu) , \\ \sigma_{P_{c,\ell}}^2 &= (2\ell+1)^2
  \, \frac{1}{2 \, N_m} \, \frac{V}{V_{w,c}} \, \int_0^1 d\mu \,
  \left[ P_c(k,\mu)^2 + \left( P_1(k,\mu) + \frac{V \sigma_1^2}{f_1^2}
    \right) \, \left( P_2(k,\mu) + \frac{V \sigma_2^2}{f_2^2} \right)
    \right] \, L_\ell^2(\mu) .
\label{eqpkmulterr}
\end{split}
\end{equation}
We note that this formulation neglects the changing line-of-sight
direction across the survey geometry.  \citet{Blake18} provide more
exact expressions for the covariance of the auto- and cross-power
spectrum multipoles, which we do not reproduce here.

\section{Simulation test}
\label{secsim}

We tested our auto- and cross-power spectrum models using a simulation
representative of a future overlapping galaxy and HI intensity mapping
dataset.  We focus on this combination of datasets because an
intensity mapping survey incorporates several observational effects
treated in Section \ref{secmodel} -- including angular/radial
pixelization, a telescope beam and pixel noise -- whereas a galaxy
survey represents a comparison sample independent of these effects.

\begin{figure}
\begin{center}
\includegraphics[width=\columnwidth]{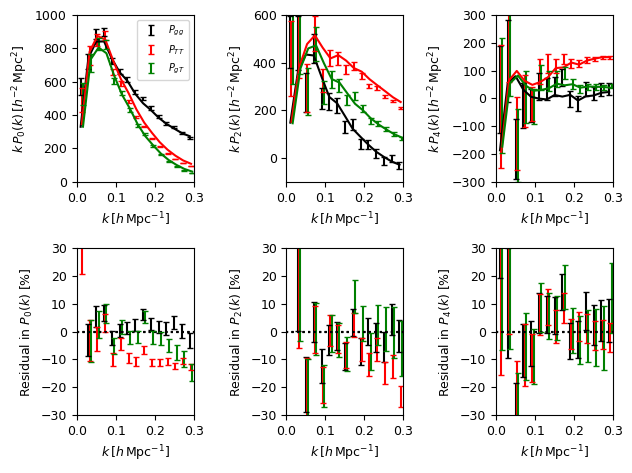}
\caption{The auto-power spectrum multipoles of the simulated galaxy
  survey $P_{gg}$ (black error bars) and intensity-mapping survey
  $P_{TT}$ (red errors), and the cross-power spectrum multipoles
  $P_{gT}$ (green errors).  The left-hand, middle and right-hand upper
  panels show the monopole ($P_0$), quadrupole ($P_2$) and
  hexadecapole ($P_4$), respectively, and the solid lines display the
  computed models in each case.  The power spectra are scaled by a
  factor of $k$ for clarity of presentation.  The lower panels display
  the residual between the measured power spectra and models,
  expressed as a percentage of the corresponding monopole power in
  each case.}
\label{figpkpole}
\end{center}
\end{figure}

Our mock dataset was built from the $z=0$ dark matter distribution of
the GiggleZ Simulation \citep{Poole15}, an N-body simulation
consisting of $2160^3$ particles evolving under gravity in a periodic
box of side $1 \, h^{-1}$ Gpc.  The initial conditions of the
simulation were generated using a fiducial flat $\Lambda$CDM
cosmological model based on the Wilkinson Microwave Anisotropy Probe
(WMAP) 5-year results \citep{Komatsu09}, with matter density $\Omega_m
= 0.273$, baryon density $\Omega_b = 0.0456$, Hubble parameter $h =
0.705$, normalization $\sigma_8 = 0.812$ and spectral index $n_s =
0.96$.  We applied the following series of steps to convert the
particle distribution into mock galaxy and intensity mapping datasets,
whose auto- and cross-correlations could be analysed using the above
theory:
\begin{enumerate}
\item We subsampled the dark matter particle distribution with a
  number density $n_p = 10^{-3} \, h^3$ Mpc$^{-3}$ within a survey
  cone defined by right ascension range $165^\circ < {\rm R.A.} <
  195^\circ$, declination range $-15^\circ < {\rm Dec.} < 15^\circ$
  and redshift range $0.3 < z < 0.7$, using the simulation fiducial
  cosmology.  This cone has a closest-fitting Fourier cuboid of volume
  $V = 0.84 \times 10^9 \, h^{-3}$ Mpc$^3$, of which a fraction $V_w/V
  = 0.54$ is observed.
\item We converted the co-moving co-ordinates of the particles into
  redshift-space positions with respect to the observer at $z = 0$,
  using the components of the particle peculiar velocities along the
  line-of-sight, whose direction varies across the cone in a
  curved-sky geometry.
\item We split the particles inside the survey cone into two random
  subsamples, which respectively formed the overlapping galaxy and
  intensity-mapping datasets, with a selection function $W_{\rm
    cone}(\vx)$ which is constant inside the cone, and zero outside
  the cone.
\item We binned the galaxy particles in a $128^3$ FFT grid, where each
  FFT grid cell has side length $\sim 7 \, h^{-1}$ Mpc, volume $\Delta
  V_{\rm FFT} = 399.7 \, h^{-3}$ Mpc$^3$, and associated Nyquist
  frequency in each dimension $k_N \sim 0.4 \, h$ Mpc$^{-1}$.
\item We binned the intensity-mapping particles in a spherical grid of
  {\tt HEALPix} cells with $N_{\rm side} = 128$ and redshift bins of
  width $\Delta z = 0.0025$.  At the centre of the survey cone, an
  angular pixel subtended $\sim 10.7 \, h^{-1}$ Mpc and a redshift
  channel extended $\sim 5.8 \, h^{-1}$ Mpc.  The angular footprint of
  the survey cone was covered by 4026 {\tt HEALPix} pixels.  We
  normalized the gridded intensity map such that pixels within the
  survey cone had a mean value of unity, $\langle f(\vx) \rangle = 1$.
\item We added uncorrelated noise to each spherical pixel $i$ of the
  simulated intensity map, drawn from a Gaussian distribution of zero
  mean and standard deviation $\sigma_i$.  We chose to create uniform
  noise $\sigma^2(\vx)$ across the survey cone by varying $\sigma_i^2$
  with the volume of each pixel $\Delta V_i$ in accordance with
  Equation \ref{eqnoisepix}, such that $\sigma_i = \sigma_{\rm fid}
  \times \sqrt{\Delta V_{\rm fid}/\Delta V_i}$, where we chose
  $\sigma_{\rm fid} = 1$ and $\Delta V_{\rm fid} = \Delta V_{\rm
    FFT}$.  From Equation \ref{eqnoisepix}, the uniform noise value is
  hence $\sigma_0^2 = \sigma_{\rm fid}^2 \, \Delta V_{\rm fid}/V$.
\item We smoothed each redshift slice of the intensity-mapping dataset
  with a Gaussian beam of standard deviation $\sigma_\theta =
  0.25^\circ$, using the {\tt HEALPix} function {\tt
    sphtfunc.smoothing}.
\item We binned the smoothed, noisy intensity-mapping dataset
  discretized in spherical pixels onto the same cubic FFT grid as the
  galaxy dataset.  We performed this step using a Monte Carlo
  algorithm, in which we generated a large number ($\sim 10^8$) of
  random points across the survey cone, and binned the random points
  in both the spherical pixels and the FFT pixels.  We then used the
  spherical binning to transfer the values of the intensity-mapping
  dataset in the spherical pixels to the random points, and averaged
  these values on the FFT grid.
\item We computed the fluctuation fields of the gridded galaxy and
  intensity-mapping datasets as $\delta(\vx) = f(\vx) - \langle f(\vx)
  \rangle$, and then estimated their auto- and cross-power spectrum
  multipoles $\lbrace P_{gg}, P_{gT}, P_{TT} \rbrace$ using Equations
  \ref{eqpkmultest} and \ref{eqpkcrossmultest}.  Since the datasets
  are uniformly distributed within the survey cone, we assumed weights
  $w(\vx) = 1$ within the cone, and set $w(\vx) = 0$ outside the cone.
  We binned our power spectrum estimates in 15 Fourier bins of width
  $\Delta k = 0.02 \, h$ Mpc$^{-1}$ in the range $0 < k < 0.3 \, h$
  Mpc$^{-1}$.
\item We subtracted Poisson noise from the galaxy power spectrum, but
  do not subtract the noise power from the intensity-mapping power
  spectrum.
\item We assigned errors to the measured multipole power spectra using
  Equation \ref{eqpkmulterr}.
\end{enumerate}
We note that the intensity mapping component of this simulated dataset
lies roughly an order of magnitude beyond the precision of current
observations, in terms of cosmological volume and noise.  We computed
the model power spectra to compare with these measurements as follows:
\begin{enumerate}
\item We generated the underlying power spectrum $P(\vk)$ of the
  fields using the RSD power spectrum of Equation \ref{eqpkrsd}.  We
  used a non-linear matter power spectrum $P_m(k)$ generated using
  {\tt CAMB} \citep{Lewis00} and {\tt halofit}
  \citep{Smith03,Takahashi12}, and adopted parameters $b_g = b_{HI} =
  1$ (given these are dark matter particles), $f = \Omega_m^{0.55} =
  0.49$ and $\sigma_v = 400$ km s$^{-1}$, which produced a good match
  to the redshift-space power spectrum of the original particle
  distribution in the full simulation cube.
\item We computed the noise component of the auto-power spectra using
  Equation \ref{eqpknoise}, $P_{\rm noise} = V \sigma_0^2 / f_0^2$.
  For both the galaxy dataset and intensity map, we included the
  contribution of the Poisson noise resulting from the discreteness of
  the original particle distribution.  This case corresponds to
  $\sigma_0^2 = f_0 = V n_p/2$.  For the intensity map, we also
  included the additional noise, with $\sigma_0^2 = \sigma_{\rm fid}^2
  \, \Delta V_{\rm fid}/V$ and $f_0 = 1$ as above.
\item For the intensity-mapping auto-power spectrum and the
  cross-power spectrum, we included a damping function, $P(\vk)
  \rightarrow P(\vk) \, D^2(\vk)$ to model the smoothing and
  pixelization.  The damping function for the auto-power spectrum of
  the intensity map is given by Equation \ref{eqdamping}, $D^2(\vk) =
  \frac{1}{V} \int d^3\vx \, |\tilde{B}(\vk,\vx)|^2$, where
  $\tilde{B}$ combines the effects of the telescope beam and spherical
  pixelization such that:
\begin{equation}
  \tilde{B}(\vk,\vx) = \tilde{B}_{\rm beam} \, \tilde{B}_{\rm chan} \,
  \tilde{B}_{\rm ang} = e^{-k_\perp^2 \, |\vx|^2 \, \sigma_\theta^2/2}
  \times \frac{\sin{(k_\parallel s_\parallel(\vx)/2)}}{k_\parallel
    s_\parallel(\vx)/2} \times W_{\rm ang}(k_\perp |\vx|) ,
\end{equation}
where $(k_\perp,k_\parallel) = (k\sqrt{1-\mu^2},k\mu)$ are the
components of $\vk$ perpendicular and parallel to the line-of-sight,
$|\vx|$ is the distance of each position from the observer,
$\sigma_\theta = 0.25^\circ$ is the Gaussian telescope beam,
$s_\parallel(\vx) = c \, \Delta z/H(z)$ is the spatial width of the
redshift bin at redshift $z$, and $W_{\rm ang}(\ell)$ is the multipole
pixel window function of the {\tt HEALPix} pixelization for $N_{\rm
  side} = 128$.  For the cross-power spectrum, we evaluated $D^2(\vk)
= \frac{1}{V} \int d^3\vx \, \tilde{B}_{\rm beam} \, |\tilde{B}_{\rm
  chan}|^2 \, |\tilde{B}_{\rm ang}|^2$, using only one power of the
telescope beam since only the intensity map is smoothed, but retaining
two powers of the pixelization for the reasons discussed at the end of
Section \ref{secpix}.  The relative contributions of these different
smoothing terms to the overall power spectrum damping are illustrated
by Figure \ref{figpkcorr}.

\begin{figure}
\begin{center}
\includegraphics[width=10cm]{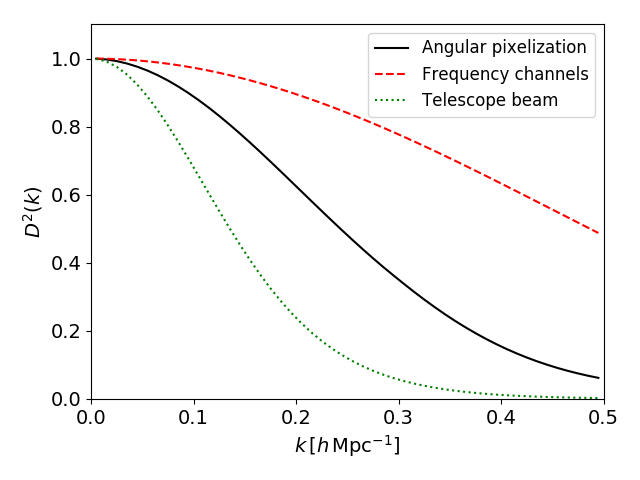}
\caption{The damping factor $D^2(\vk)$ defined by Equation
  \ref{eqdamping} for the simulation tests described in Section
  \ref{secsim}.  We show cases corresponding to the angular
  pixelization (black solid line) where $(k_\perp=k, k_\parallel=0)$,
  the radial frequency channels (red dashed line) where $(k_\perp=0,
  k_\parallel=k)$ and the telescope beam (dotted green line) where
  $(k_\perp=k, k_\parallel=0)$.}
\label{figpkcorr}
\end{center}
\end{figure}

\item We convolved the damped, model power spectra with the window
  function of the survey cones, $P(\vk) \rightarrow P \star |W_{\rm
    cone}|^2$.
\item To allow for the discretization onto the FFT grid, we summed the
  resulting power spectra over modes $\vk_\vn = \vk + 2k_N\vn$ using
  Equation \ref{eqdisc}, taking a $3^3$ grid of $\vn = \lbrace -1,0,1
  \rbrace$.
\item We averaged the model power spectra in the same Fourier bins as
  the measurements.
\end{enumerate}
The auto- and cross-power spectrum multipole measurements and models
for this test case, along with the residuals, are displayed in Figure
\ref{figpkpole}.  There is good general agreement between the models
and mock observations.  The most significant deviation occurs for the
monopole of the intensity power spectrum $P_{TT}$, whose measured
amplitude lies around $10\%$ below the model.  We attribute this
offset to the approximations implemented when deriving the damping of
the model due to spherical pixelization\footnote{We found closer
  agreement between the model and simulations in a flat-sky case with
  regular pixelization; we leave this issue for future work.}, as
described in Section \ref{secsmooth}.  The $P_{gT}$ monopole and
$P_{TT}$ quadrupole also show some deviations for scales $k > 0.2 \,
h$ Mpc$^{-1}$.  Excepting the $P_{TT}$ monopole, all statistics and
multipoles produce a satisfactory value of the $\chi^2$ statistic with
$\chi^2/{\rm dof} \sim 1$ for scales $k < 0.2 \, h$ Mpc$^{-1}$.

Current cross-correlation analyses of radio intensity mapping and
galaxy surveys have produced $\sim 3$-$\sigma$ detections of the
cross-power spectrum amplitude, and corresponding constraints on the
neutral hydrogen density at intermediate redshifts
\citep{Chang10,Masui13,Wolz17,Anderson18}.  Analysis of the auto-power
spectrum of intensity maps is currently severely limited by imperfect
foreground subtraction.  Hence, we conclude that our model is likely
to remain sufficient for the analysis of near-future intensity mapping
datasets, which will focus on cross-correlation measurements.

\section{Summary}
\label{secconc}

In this paper we have provided a general framework connecting the
measured 2-point auto- and cross-correlations of fluctuation fields to
their underlying cosmological power spectra, in the presence of a
variety of observational effects.  Our framework can be applied to the
analysis of galaxy spectroscopic redshift surveys, datasets with
accurate photometric redshifts, radio intensity-mapping surveys, or
other 3D cosmological maps.

The observational effects we considered are the variation with
position of the background level of the field, measurement noise, the
smoothing and discretization of the field, and the changing
line-of-sight direction.  We extended previous literature by deriving
that if a field is smoothed by a position-dependent kernel,
$B(\vs,\vx)$, where $\vs$ is the kernel separation with respect to
position $\vx$, then the power spectrum is damped by the
volume-average of the Fourier transform of the kernel at each
position, $\frac{1}{V} \int d^3\vx \, |\tilde{B}(\vk,\vx)|^2$, under
the approximation that $\tilde{B}(\vk,\vx)$ varies slowly with $\vx$.
We applied this result to the cases of averaging a field in irregular
cells, applying noise in these cells, a telescope beam, redshift
errors, and binning data in frequency channels and angular pixels.

We reviewed the direct estimators of the auto- and cross-power
spectra, their multipoles, and the variance in these statistics.  We
extended the results of \citet{FKP} to present optimal weights for
measuring the auto- and cross-power spectra of general cosmological
fluctuation fields, with application to galaxy surveys and intensity
maps.  FKP weights for individual tracers do not in general provide
the optimal weights when measuring the cross-power spectrum.

We validated our model by comparison with the power spectrum
multipoles of a mock galaxy and intensity-mapping dataset drawn from
an N-body simulation, including several of these observational
effects.  The intensity mapping component of this simulated dataset
lies roughly an order of magnitude in precision beyond current
observations.  The model is effective in reproducing the measured
statistics, excepting a $\sim 10\%$ residual in the monopole of the
intensity auto-power spectrum, $P_{TT}$.  However, given that current
analyses of $P_{TT}$ are limited by imperfect foreground subtraction,
our model is likely to remain sufficient for the analysis of
near-future intensity-mapping datasets.  We note that a number of
other observational effects, such as fibre collisions, selection
function systematics, radio foregrounds and photometric redshift
outliers, are not considered in this study but may be relevant for the
analysis of real data.

We hope that this study has provided a set of recipes and derivations
which will be useful for modelling and studying the Fourier-space
statistics of cosmological fluctuation fields in observed and
simulated datasets.  Accompanying power spectrum code for producing
our mock dataset and evaluating the measurements and models is
available at \url{https://github.com/cblakeastro/intensitypower}.

\section*{Acknowledgements}

We are grateful to the anonymous referee for their thorough reading of
the paper and numerous constructive suggestions.  We thank Laura Wolz,
Eva-Maria Mueller, Rossana Ruggeri, Alkistis Pourtsidou, Steve
Cunnington, David Bacon and Fei Qin for valuable discussions during
the development of this project, and Cullan Howlett for useful
comments on a draft of this paper.  The GiggleZ N-body simulation used
in this work was originally generated and shared by Greg Poole
\citep{Poole15}.  Some of the results in this paper have been derived
using the HEALPix package \citep{Healpix05}.  We have used {\tt
  matplotlib} \citep{Hunter07} for the generation of scientific plots,
and this research also made use of {\tt astropy}, a
community-developed core Python package for Astronomy
\citep{Astropy13}.

\bibliographystyle{mnras}
\bibliography{intensity_power}

\appendix

\bsp
\label{lastpage}
\end{document}